\documentclass[aps,amssymb,twocolumn]{revtex4}
\usepackage{graphicx}

\begin{document}
\title{Superconducting energy gap in MgB$_{2}$ film observed by infrared
reflectance.}
\author{A. Pimenov$^{1}$, A. Loidl$^{1}$, and S. I. Krasnosvobodtsev$^{2}$}
\address{$^{1}$Experimentalphysik V, EKM, Universit\"{a}t Augsburg, 86135 Augsburg,
Germany \\ $^{2}$P. N. Lebedev Physics Institute, Russian Academy
of Sciences, 117924 Moscow, Russia}
\date{January 9, 2002}

\begin{abstract}
Far-infrared reflectance of a MgB$_{2}$ film has been measured by
Fourier-transform spectroscopy for frequencies 10\,cm$^{-1}<\nu
<4000$\,cm$^{-1}$ above and below the superconducting transition.
The data provide clear experimental evidence for the onset of a
superconducting gap at 24\,cm$^{-1}$ at $T=5$\,K. On increasing
temperature the gap energy increases, contrary to what is expected
in isotropic BCS superconductors. The small zero-temperature gap
value and its unconventional increases on increasing temperature
can only be explained by a highly anisotropic or multiple gap
function.
\end{abstract}

\maketitle
%\pacs{74.25.Gz, 74.76.-w, 74.25.-q}
%\begin{multicols}{2}

The discovery of superconductivity in MgB$_2$ \cite{nagamatsu}
raised the question about the nature of the energy gap in this
compound. Among different methods to observe and measure the
superconducting energy gap, infrared spectroscopy remains a very
promising and powerful method. Indeed, first experiments on
grazing infrared reflectivity in MgB$_2$ \cite{gorshunov} showed
features of the experimental spectra, which may be attributed to
the superconducting gap. The analysis of the data revealed a gap
value in the range $2 \Delta \simeq 3-4 $\,meV. This value is
rather small compared to the BCS estimate $2 \Delta \approx
12$\,meV and has been attributed to the minimum of the gap
distribution due to anisotropy.

The infrared-transmission experiments by Jung. \textit{et al.}
\cite{jung} carried out on MgB$_2$ thin films revealed a
characteristic peak, which could be connected to the energy gap
and resembled the predictions of the weak-coupling BCS-theory.
These experiments were similar to the classical experiments of
Ginsberg and Tinkham on lead, tin and indium \cite{ginsberg}.
Applying the Kramers-Kronig analysis to the thin-film
transmission data of these compounds, a superconducting gap in the
real part of the complex conductivity, $\sigma^* (\omega)=
\sigma_1 + i\sigma_2$ could be directly observed. In experiments
with MgB$_2$, Jung. \textit{et al.} \cite{jung} restricted the
data analysis to the comparison with predictions of the BCS-theory
which gave an estimate of the energy gap, $2\Delta_0 \simeq
5.2$\,meV. Later on, terahertz time-domain spectroscopy has been
applied to the transmission of thin MgB$_2$ films \cite{kaindl}.
Due to phase-sensitivity of the time-domain spectroscopy, the
complex conductivity of MgB$_2$ could be extracted from the
spectra without application of the Kramers-Kronig analysis. As for
classical superconductors, the frequency dependence of $\sigma_1$
revealed a characteristic minimum in the frequency dependence,
which could be attributed to the superconducting gap, $2\Delta
\simeq 5$\,meV.

While there is already a large scatter of gap values as determined
from optical experiments, the gap energies differ even  more when
values deduced from different experimental setups are compared.
Techniques like tunneling spectroscopy \cite{karapetrov,giubileo},
photoemission \cite{takahashi,tsuda} or specific heat
\cite{bouquet} reveal significantly different values of the energy
gap ranging between $2\Delta=3-10$\,meV . These discrepancies
together with the observed deviations from the BCS temperature
dependence have been partly explained by imperfections of the
sample surface, but alternative explanations based e.g. on
multiple gaps \cite{tsuda,bascones} or  anisotropy \cite{haas}
have also been proposed. For a recent discussion see Ref.
\cite{buzea} and references therein. Different values, observed by
various experiments may be partly explained by the preferential
sensitivity of particular experiments to different portions of the
gap distribution. Probably, optical conductivity methods tend to
yield rather the lower limit of the gap distribution. The
above-mentioned infrared experiments \cite{gorshunov,jung,kaindl}
revealed the $2\Delta$ values between 3 and 5 meV, i.e
substantially below the BCS estimate.

Previously, we have investigated thin films of MgB$_2$ using the
method of quasioptical spectroscopy in the frequency range 4
cm$^{-1}<\nu <30$ cm$^{-1}$ \cite{pronin}. The highest frequency
of this experiment was too low to make distinct conclusions about
the spectral gap in $\sigma_1 (\omega)$. Therefore, we
investigated the same sample at infrared frequencies using the
technique of Fourier-Transform spectroscopy. The infrared
spectroscopy of thin superconducting films reveals distinct
advantages compared to the spectroscopy on bulk samples
\cite{ferro}. The reflectance of bulk superconducting samples
rapidly approaches unity both in the normal and in the
superconducting state especially at low frequencies. This implies
a number of experimental difficulties investigating the changes in
reflectance at the superconducting transition. In contrast, the
reflectance of thin films can be substantially reduced (compared
to unity), which strongly improves the accuracy of the data. In
addition, the technique of quasioptical transmission spectroscopy
\cite{kozlov} can be applied to the same films at lower
frequencies. The transmission technique provides an independent
measurement of the complex conductivity at submillimeter
frequencies and the reflectance of the sample can be calculated
from these data, which strongly expands the low-frequency limit
of the available spectrum and substantially improves the quality
of the Kramers-Kronig analysis. The combination of these two
techniques has been applied previously to thin YBaCuO films
\cite{ferro} and showed the reliability and the advantage of the
described procedure compared to conventional far-infrared
analyses.

The MgB$_{2}$ film  has been grown by two-beam laser ablation on
a $[1\bar{1}02]$ sapphire substrate. Magnetization measurements
 have indicated a sharp superconducting transition at 32 K with
a width of 1 K. The details of the growth process will be given
elsewhere \cite{prep}. For further details of the  film
characterization see Ref. \cite{pronin}. In the frequency range
$10<\nu <4000$\,cm$^{-1}$ reflectivity measurements were
performed using a Bruker IFS-113v Fourier-transform spectrometer.
Different sources, beam splitters and optical windows allowed to
cover the complete frequency range. A pumped liquid-He bolometer
has been employed below 60\,cm$^{-1}$. In addition, the
reflectance in the frequency range $4<\nu <30$\,cm$^{-1}$ has
been calculated using the complex conductivity data of the same
sample, obtained by the submillimeter transmission \cite{pronin}.
Details of this technique can be found elsewhere \cite{kozlov}.

\begin{figure}[tb]
\centering
\includegraphics[width=7cm,clip]{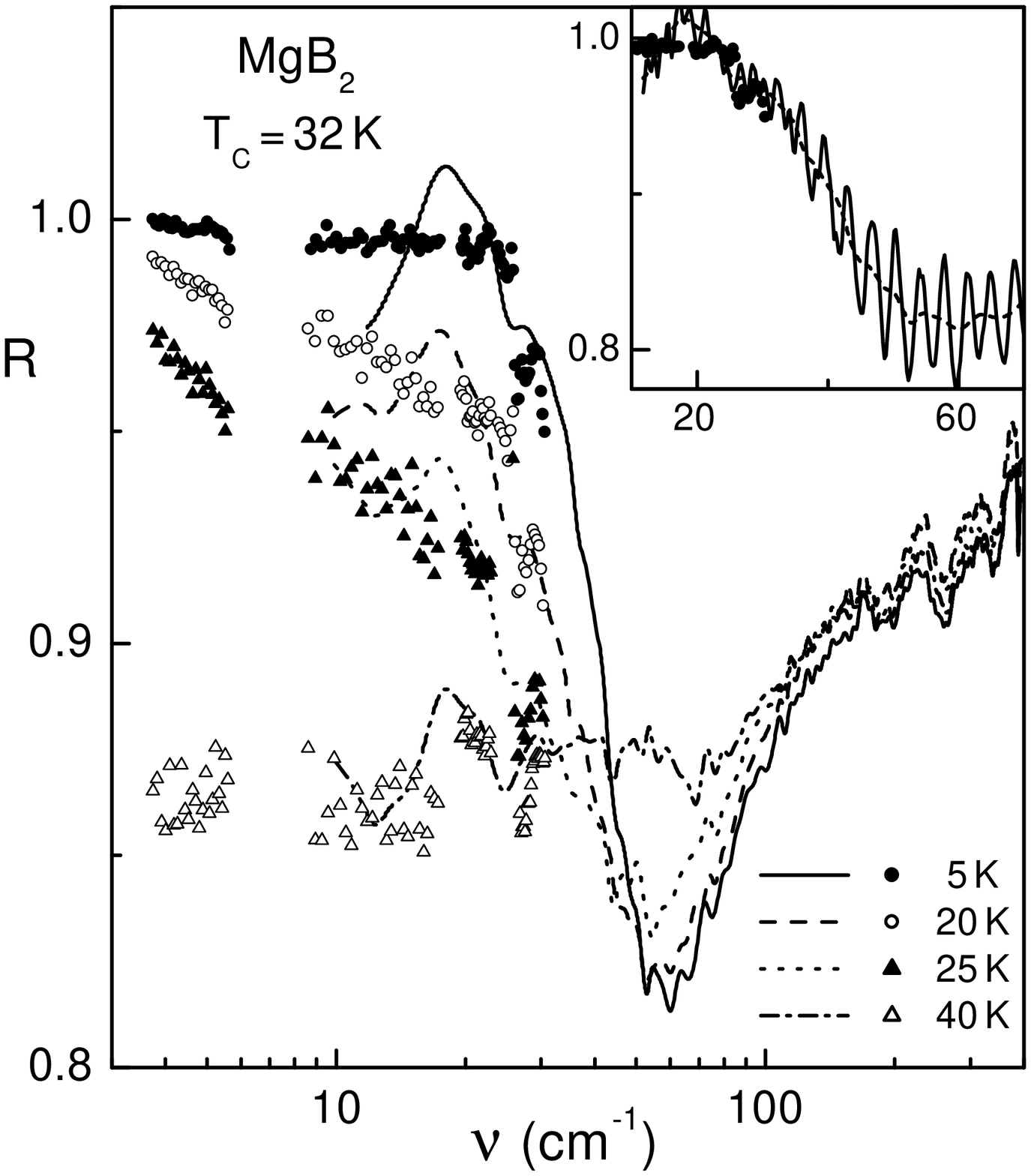}
\vspace{0.2cm} \caption{Far-infrared and submillimeter reflectance
of MgB$_{2}$ film on an Al$_{2}$O$_{3}$ substrate. Lines: directly
measured reflectance, symbols: data as calculated from the complex
conductivity in the submillimeter range \protect\cite{pronin}. The
inset demonstrates the result of the averaging procedure of the
interference pattern at $T=5$\,K.} \label{frefl}
\end{figure}

Fig.\ \ref{frefl} shows the reflectance of the MgB$_2$ film for
different temperatures. The most striking feature of the data is
the change in slope of the reflectance around 24 cm$^{-1}$. The
reflectance below this frequency is close to one at $T=5$\,K and
drops rapidly above 24 cm$^{-1}$. It is important to note that
this change in slope is seen in the calculated submillimeter data
(symbols) as well. Therefore, we interpret this feature as a
direct observation of the spectral gap in MgB$_2$. The abrupt
change in slope around 24 cm$^{-1}$ clearly indicates the onset of
the electromagnetic absorption above this frequency. It therefore
resembles the classical experiments on reflectance in conventional
superconductors \cite{richards}, in which the onset absorption
across the gap has been observed. The onset of the absorption in
MgB$_2$ corresponds to a superconducting gap $2\Delta\simeq
3$\,meV, which is a factor of four smaller than the BCS-estimate
and is in the lower limit of the gap values, observed in other
experiments ($3-10$\,meV). As mentioned above, the energy gaps,
estimated from the infrared \cite{gorshunov,jung,kaindl} and
microwave \cite{klein} experiments, all reveal quite small values
lying in the range $3-5$\,meV. We recall further that according to
 first-principles calculations \cite{kortus}, the Fermi-surface
of MgB$_2$ comprises four sheets, each having a distinct energy
gap \cite{bascones}. Indeed, recent tunneling spectroscopy
experiments \cite{giubileo} revealed a two different energy gaps
in MgB$_2$. It is reasonable to suggest that the experiments which
measure the complex conductivity (e.g. microwave or infrared
spectroscopy) are sensitive to the lowest gap value. The
spectroscopic features due to other gaps are smoothed because they
overlap with the absorption across the smallest energy gap. This
effect is a possible reason for a relatively low frequency of the
"knee" in reflectance observed in Fig.\ \ref{frefl}.

The reflectivity of a thin metallic film on a dielectric substrate
may be obtained from the Maxwell equations and takes the form
\cite{heavens}:
\begin{equation}
r=\frac{r_{0f}+r_{fs}\exp ({4\pi }in_{f}d{ /\lambda
)}}{1+r_{0f}r_{fs}\exp ({4\pi }in_{f}d{/\lambda )}}\quad ,
\label{erefl}
\end{equation}
with $r_{0f}=(1-n_f)/(1+n_f)$ and $r_{fs}=(n_f-n_s)/(n_f+n_s)$
being the Fresnel reflection coefficients at the air-film
$(r_{0f})$ and film-substrate $(r_{fs})$ interface. Here $n_f =
(i\sigma^* /\varepsilon_0 \omega^*)^{1/2}$ and $n_s$ are the
complex refractive indices of the film and substrate,
respectively, $\lambda $ is the radiation wavelength, $d$ is the
film thickness, $\omega= 2\pi \nu$ is the angular frequency,
$\sigma^*=\sigma_1+i\sigma_2$ is the complex conductivity of the
film, and $\varepsilon_0$ is the permittivity of free space.

Eq. (\ref{erefl}) neglects reflections from the opposite side of
the substrate. It has to be noted that especially at low
frequencies, interferences within the substrate are observed in
the spectra (inset of Fig. \ref{frefl}). In that case the full
expression taking into account the finite substrate thickness is
more appropriate to calculate the conductivity. However, attempts
to carry out the calculation using the exact expression instead of
Eq. (\ref{erefl}) did not improve the quality of the data. A
possible reason for this fact is the insufficient resolution of
the interferences. Instead, the interference fringes in the
reflectivity spectra have been removed by averaging over the
frequency range of one oscillation. The results of the averaging
procedure are demonstrated in the inset of Fig. \ref{frefl} and do
not change the overall frequency dependence of the reflectance.

If the film thickness is smaller than the penetration depth
($\left| n_{f}\right| d\gg \lambda $) and if $\left| n_{f}\right|
\gg \left| n_{s}\right| $, Eq. (\ref{erefl}) is simplified to :
\begin{equation}
r \approx \frac{1-\sigma ^{*}dZ_{0}-n_{s}}{1+\sigma
^{*}dZ_{0}+n_{s}} \label{erefl1} ,
\end{equation}
where $\sigma ^{*}=-in_{f}^{2}\varepsilon _{0}\omega $ is the
complex conductivity of the film  and
$Z_{0}=\sqrt{\mu_0/\varepsilon_0} \simeq 377\,\Omega$ is the
impedance of free space. Eq. (\ref{erefl1}) is advantageous due
to its simplicity and provides a good approximation of the
reflectivity at submillimeter frequencies. For higher frequencies
Eq. (\ref{erefl}) is more appropriate.

The experimentally obtained reflectance $R=|r|^2$ does not
provide enough information to calculate the film conductivity
because of the unknown phase shift. Therefore, a Kramers-Kronig
analysis has to be applied to the reflectance in order to obtain
the phase shift. The condition of the applicability of this kind
of analysis is the analyticity of $\log |r|$ in the upper half
plane \cite{cardona}. The absence of the poles in $\log |r|$ is
ensured by the relation $|r| \le 1$. The absence of zeros in
$|r|$ (i.e. branch points in $\log |r|$) is demonstrated in Fig.
\ref{frefl} and follows from the analysis of the numerator of Eq.
(\ref{erefl}): the term $r_{0f}$ gives the reflectivity of the
metal-vacuum boundary and  therefore is close to (-1), the term
$r_{fs}\exp ({4\pi }in_{f}d{ /\lambda )}$ is suppressed due to
the presence of the exponent.

The general problem in the application of the Kramers-Kronig
analysis is the necessity to extend the reflectance spectrum to
low and to high frequencies. The high-frequency extrapolation up
to 22000 cm$^{-1}$ has been taken from the data of Tu \textit{et
al.} \cite{tu}. The low-frequency extrapolation, which is the
most important for our data analysis,
 has been obtained from the complex
conductivity data of the same film, measured by quasioptical
transmission spectroscopy in the range $4<\nu<30$\,cm$^{-1}$
\cite{pronin}. Below 4 cm$^{-1}$ the extrapolations $(1-r) \propto
\omega$ and $r=const$ have been used in the superconducting and
normal conducting state, respectively. These extrapolations follow
from Eq. (\ref{erefl1}) and from the low-frequency limits
$\sigma^* \approx i\sigma_2\propto1/\omega$ for a superconductor
and $\sigma^* \approx \sigma_1$ for a metal. As demonstrated in
Fig. \ref{frefl}, the calculated submillimeter data (symbols)
overlap and agrees well with the measured infrared reflectance
(lines) without any arbitrary shifts.

\begin{figure}[]
\centering
\includegraphics[width=7cm,clip]{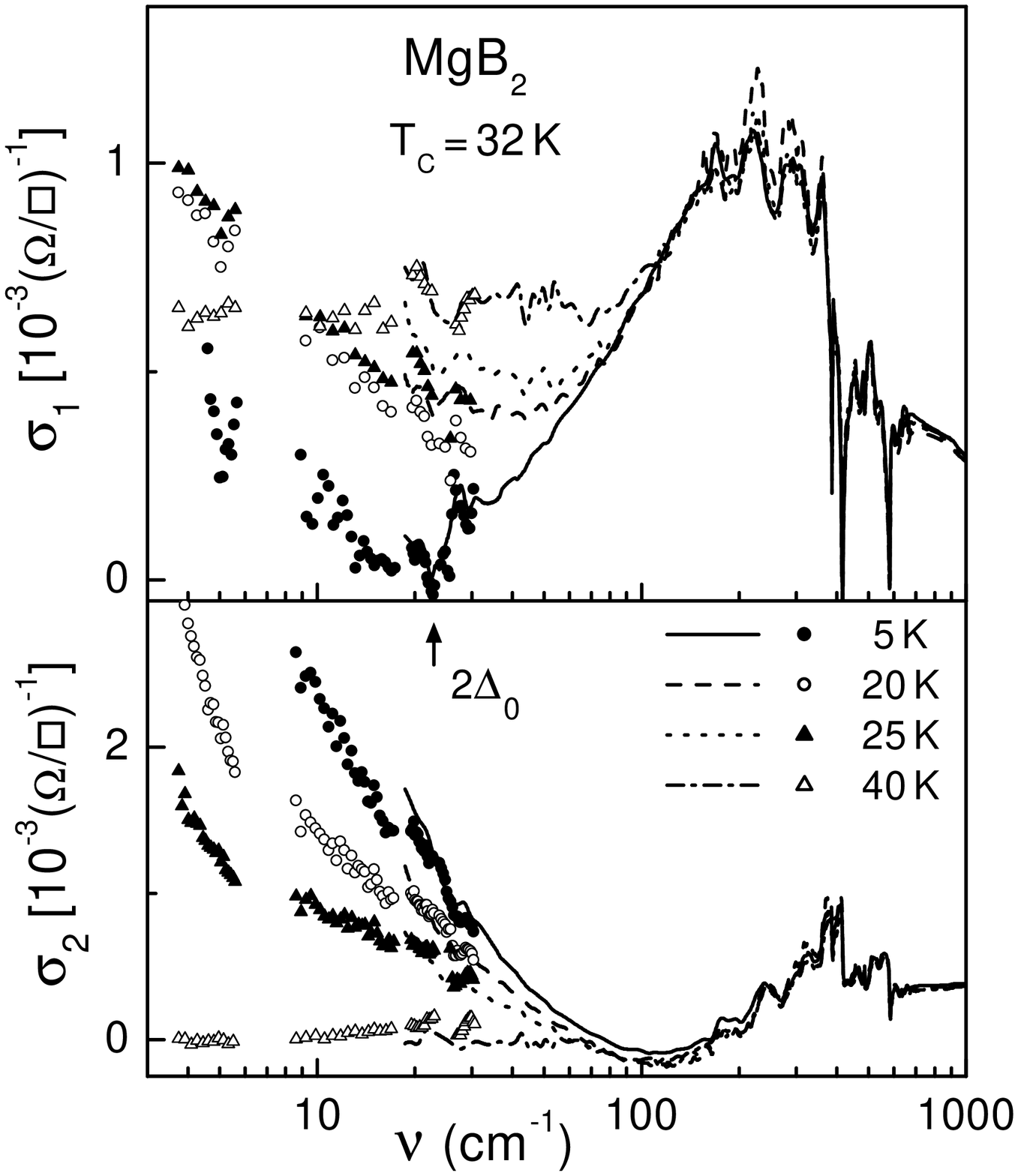}
\vspace{0.2cm} \caption{Complex conductivity of a  MgB$_{2} $ film
at infrared frequencies. Upper panel: real part $\sigma _{1}$.
Lower panel: imaginary part $\sigma _{2}$. Lines have been
obtained via the Kramers-Kronig analysis of the reflectance data.
Symbols represent the conductivity, which was directly measured by
the transmittance technique \protect\cite{pronin}. Sharp peak-like
structures between 400 and 700\,cm$^{-1}$ is due to residual
influence of phonon contribution of the substrate.} \label{fcond}
\end{figure}

The complex conductivity of the MgB$_2$ film has been calculated
from the reflectance data of Fig.\ \ref{frefl} and solving Eq.\
(\ref{erefl}). The complex refractive index of the Al$_2$O$_3$
substrate has been obtained in a separate experiment and agrees
well with literature data \cite{barker}. The reflectivity phase
shift of the substrate-film system has been obtained by carrying
out the Kramers-Kronig analysis as described above. The results
are presented in Fig. \ref{fcond}. The real part of the
conductivity (upper panel) is frequency independent in the normal
conducting state ($T=40$\,K) and below 100\,cm$^{-1}$. The
corresponding imaginary part $\sigma_2$ (lower panel) is roughly
zero for these frequencies. This picture corresponds well to the
low-frequency limit of the Drude-conductivity of a simple metal.
On cooling the sample into the superconducting state, $\sigma_1$
becomes suppressed between 5 and 100 cm$^{-1}$ and shows a clear
minimum, which corresponds to the superconducting gap. The
position of the minimum at $T=5$\,K corresponds well to the change
in slope in reflectance ($\nu\approx 24$\,cm$^{-1}$, Fig.
\ref{frefl}). The minimum in $\sigma_1$ shifts to \emph{higher}
frequencies as the temperature increases. This apparently
contradicts  the expectations for the temperature dependence of
the superconducting energy gap. However, as already mentioned
discussing the reflectance data of Fig.\ \ref{frefl}, MgB$_2$
reveals either an anisotropic gap \cite{haas} or multiple gaps
\cite{bascones}. At the lowest temperatures the smallest gap
dominates and the spectral features at higher frequencies due to
other gaps are masked. On increasing temperature the transition
across the smallest gap gradually becomes thermally-saturated and
the effective "weight" for different gaps is shifted to the
higher-energy region. Therefore the conductivity minimum also
shifts to higher frequencies, which explains the behavior observed
in Fig. \ref{fcond}.

The suppression of $\sigma_1$ at infrared frequencies (upper part
of Fig. \ref{fcond}) reflects the reduction of the low-frequency
spectral weight. At the same time, $\sigma_2$ rapidly increases
revealing approximately a $1/\omega$ behavior. From the slope of
$\omega^{-1}$ the spectral weight of the superconducting
condensate can be estimated and turns out to be roughly 20\%
higher than the reduction of the spectral weight in $\sigma_1$.
This possibly indicates that a substantial extrinsic contribution
is present in $\sigma_1$ in the superconducting state.

Besides the gap feature, an additional excitation is observed in
the conductivity spectra for frequencies above 100\,cm$^{-1}$
(Fig. \ref{fcond}). This feature is seen as a broad maximum
around 230\,cm$^{-1}$, which is only weakly temperature
dependent. For frequencies higher than 400\,cm$^{-1}$ the
conductivity decreases again, which possibly reflects the
high-frequency behavior of the Drude-conductivity with a
characteristic scattering rate $1/2\pi\tau \approx
400$\,cm$^{-1}$. At the same frequency a broad maximum becomes
apparent in $\sigma_2$ (lower part of Fig. \ref{fcond}), a
behavior which is expected within the Drude picture. The
scattering rate $1/2\pi\tau \approx 400$\,cm$^{-1}$ roughly
corresponds to $1/2\pi\tau \approx 150$\,cm$^{-1}$, obtained from
the low-frequency transmittance of this film \cite{pronin}. The
discrepancies most probably reflect the experimental
uncertainties. The origin of the broad maximum in $\sigma_1$ on
 top of the Drude-conductivity is unclear at present. We
cannot attribute it to  residual effects due to substrate,
because the phonon excitations in Al$_2$O$_3$ start around 400
cm$^{-1}$ and produce sharp anomalies in the spectra for
$400<\nu<700$\,cm$^{-1}$. A possible alternative explanation for
the broad maximum could be bound states of localized charge
carriers, similar to infrared maxima, observed in Zn-doped
YBa$_2$Cu$_3$O$_8$ \cite{basov}.

\begin{figure}[]
\centering
\includegraphics[width=7cm,clip]{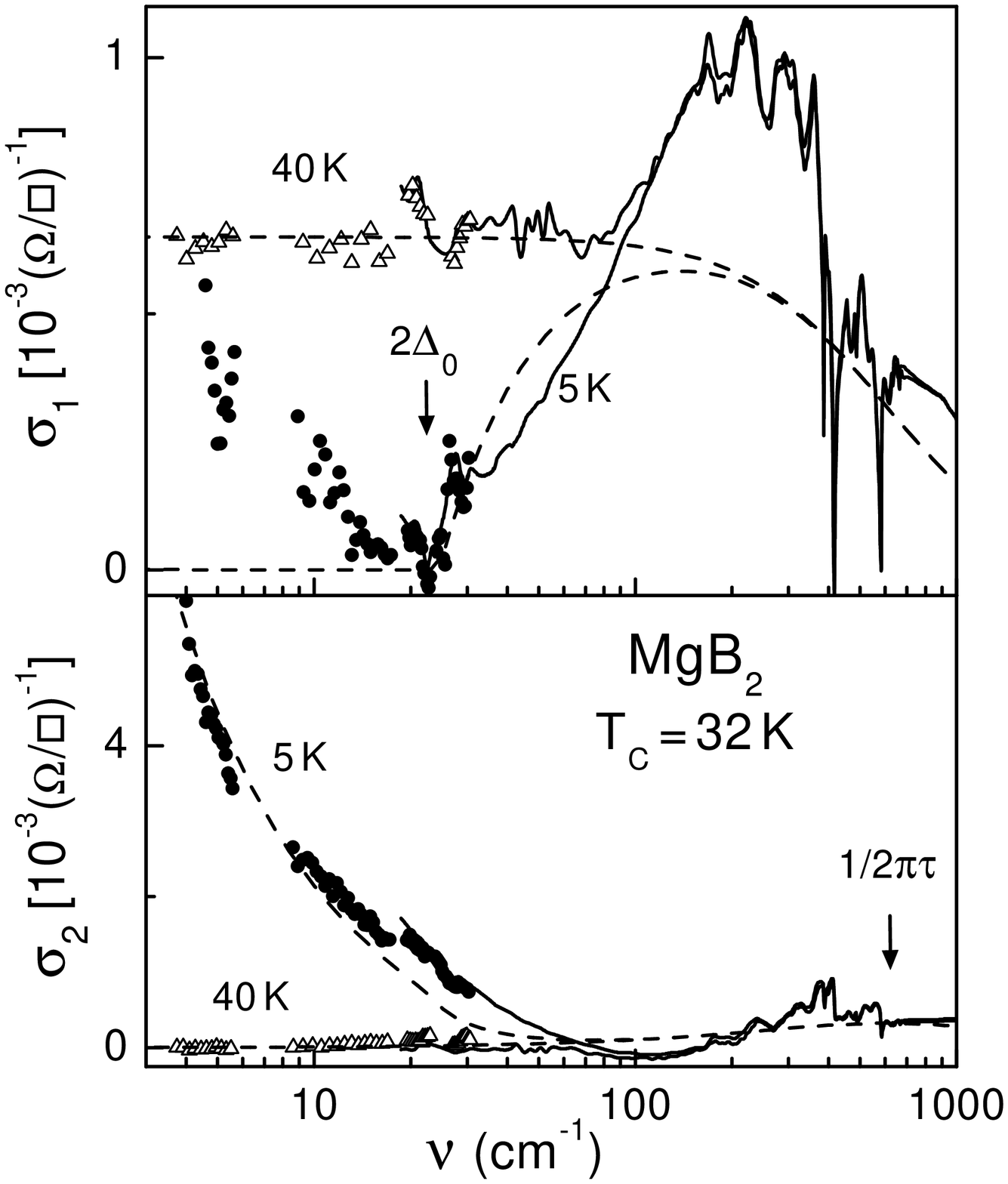}
\vspace{0.2cm} \caption{Comparison between the complex
conductivity and the predictions of the weak-coupling BCS-model.
Symbols and solid lines represent the experimental data of Fig.
\ref{fcond}. Dashed lines: BCS-theory with
$2\Delta=24$\,cm$^{-1}$, $1/2\pi\tau=600$\,cm$^{-1}$.}
\label{ftheory}
\end{figure}

Finally, Fig. \ref{ftheory} compares the experimental conductivity
with the predictions of the weak-coupling BCS-model \cite{brandt}.
The value of the superconducting gap is fixed by the
characteristic feature in $\sigma_1$. The scattering rate
$1/2\pi\tau=600$\,cm$^{-1}$ has been chosen in order to reasonably
fit the decrease of $\sigma_1$ at high frequencies and the broad
maximum in $\sigma_2$ around 400\,cm$^{-1}$. This value of the
scattering rate implies the applicability of the dirty limit to
MgB$_2$. In this limit the changes of the scattering rate have no
substantial influence on the complex conductivity below
100\,cm$^{-1}$.  However, we recall that the energy gap
$2\Delta=24$\,cm$^{-1}$ disagree with the BCS-value $2\Delta_{BCS}
\simeq 80$\,cm$^{-1}$ ($T_{\rm c}=32$\,K). In this case the
calculations based on the multi-gap conductivity might be more
appropriate. We expect that the presented estimates within the
simple model would give qualitative hints to understand the
behavior of the complex conductivity.

The imaginary part of the conductivity at $T=5$\,K is well
reproduced by the BCS-model without additional fitting parameters.
For low frequencies and in the superconducting state the real part
$\sigma_1$ is far above the model calculations, a fact that
probably implies an additional absorption in the film. Above the
gap frequency, the theoretical curve shows a slightly higher slope
compared to the experiment. This again may indicate a distribution
of the superconducting gaps in MgB$_2$.

In conclusion, we presented the far-infrared reflectance of a
thin MgB$_2$ film and combined it to the low-frequency data,
obtained by the submillimeter spectroscopy. The broadband
reflectance reveals a well-defined change in slope around $\nu
\simeq 24$\,cm$^{-1}$, which we interpret as a direct evidence of
the superconducting gap, $2\Delta\simeq3$\,meV. The complex
conductivity at infrared frequencies has been obtained applying
the Kramers-Kronig analysis to the reflectance data and solving
the equations for the substrate-film system. In the real part of
the conductivity a minimum develops in the superconducting state,
which corresponds to the gap frequency. In addition, a broad
conductivity maximum is observed at $\nu \sim 230$\,cm$^{-1}$ and
is attributed to bound states of localized charge carriers.

We thank F. Mayr, Ch. Hartinger and K. Pucher for help in
carrying out the infrared measurements, and S. Six, Ch. Schneider
and G. Hammerl for the film characterization. We acknowledge
stimulating discussion with A. V. Pronin and H.-A. Krug von Nidda.
This work was supported by the BMBF via the contract 13N6917/0 -
EKM.

%\end{multicols}

\begin{thebibliography}{99}

\bibitem{nagamatsu}  J. Nagamatsu {\em et al.}, Nature {\bf 410}, 63 (2001).

\bibitem{gorshunov}  B. Gorshunov {\em et al.}, Eur. Phys. J. B \textbf{21}, 159 (2001).

\bibitem{jung}  J. H. Jung {\em et al.}, cond-mat/0105180.

\bibitem{ginsberg}  D. M. Ginsberg and M. Tinkham, Phys. Rev. \textbf{118},
990 (1960).

\bibitem{kaindl} R. A. Kaindl \textit{et al.}, Phys. Rev. Lett. \textbf{88}, 027003 (2002).

\bibitem{karapetrov}  G. Karapetrov {\em et al.}, Phys. Rev.
Lett. \textbf{86}, 4374 (2001); G. Rubio-Bollinger, H. Suderow,
and S. Vieira, Phys. Rev. Lett. \textbf{86}, 5582 (2001); A.
Sharoni {\em et al.}, Phys. Rev. B \textbf{63}, 220508 (2001); H.
Schmidt {\em et al.}, Phys. Rev. B \textbf{63}, 220504 (2001).

\bibitem{giubileo}  F. Giubileo
\textit{et al.}, Phys. Rev. Lett. \textbf{87}, 177008 (2001); H.
Schmidt \textit{et al.}, cond-mat/0112144.

\bibitem{takahashi}  T. Takahashi {\em et al.}, Phys. Rev. Lett. \textbf{86}, 4915 (2001).

\bibitem{tsuda} S. Tsuda {\em et al.}, Phys. Rev. Lett. \textbf{87}, 177006 (2001).

\bibitem{bouquet} F. Bouquet {\em et al.}, Phys. Rev. Lett. \textbf{87}, 047001
(2001); R. K. Kremer {\em et al.}, cond-mat/0102432; Ch. W\"{a}lti
{\em et al.}, Phys. Rev. B \textbf{64}, 172515 (2001); Y. Wang
{\em et al.}, Physica C \textbf{355}, 179 (2001).

\bibitem{bascones}  E. Bascones and F. Guinea, Phys. Rev. B \textbf{64}, 214508 (2001);
S. V. Shulga \emph{et al.}, cond-mat/0103154; A. Y. Liu, I. I.
Mazin, and J. Kortus, Phys. Rev. Lett. \textbf{87}, 087005 (2001).

\bibitem{haas} S. Haas and K. Maki, Phys. Rev. B \textbf{65}, 020502 (2002).

\bibitem{buzea} C. Buzea and T. Yamashita, Supercond. Sci. Technol. \textbf{14}, R115 (2001).

\bibitem{pronin} A. V. Pronin \textit{et al.}, Phys. Rev. Lett. \textbf{87}, 097003
(2001).

\bibitem{ferro}  A. Pimenov \textit{et al.}, Ferroelectrics, \textbf{249}, 165 (2001).

\bibitem{kozlov}  G. V. Kozlov and A. A. Volkov, in {\em Millimeter and
Submillimeter Wave Spectroscopy of Solids}, edited by G.
Gr\"{u}ner (Springer, Berlin, 1998), p. 51.

\bibitem{prep}  S. I. Krasnosvobodtsev {\em et al.}, to be
published.

\bibitem{richards} P. L. Richards and M. Tinkham, Phys. Rev.
\textbf{119}, 575 (1960); S. L. Norman and D. H. Douglass, Jr.,
Phys. Rev. Lett. \textbf{18}, 1967 (1967).

\bibitem{klein} N. Klein \emph{et al.}, cond-mat/0107259; F. Manzano
\emph{et al.}, Phys. Rev. Lett. \textbf{88}, 047002 (2002).

\bibitem{kortus} J. Kortus \emph{et al.}, Phys. Rev. Lett. \textbf{86},
4656 (2001).

\bibitem{heavens}  O. S. Heavens, {\it Optical properties of thin solid films}
 (Dover Publ., New York, 1991).

\bibitem{cardona} M. Cardona, in Optical Properties of Solids, Edited
by S. Nudelman and S. S. Mitra (Plenum Press, New York, 1969), p.
137.

\bibitem{tu} J. J. Tu \textit{et al.}, cond-mat/0107349.

\bibitem{barker} A. S. Barker, Jr., Phys. Rev. \textbf{132}, 1474 (1963).

\bibitem{basov}  D. N. Basov, B. Dabrowski, and T. Timusk, Phys. Rev.
Lett. {\bf 81}, 2132 (1998).

\bibitem{brandt} W. Zimmermann \textit{et al.}, Physica C \textbf{183}, 99 (1991).


\end{thebibliography}
\end{document}